\documentstyle[12pt] {article}
\newlength{\extraspace}
\newlength{\extraspaces}
\newcommand{\be}{\begin{equation}
\addtolength{\abovedisplayskip}{\extraspaces}
\addtolength{\belowdisplayskip}{\extraspaces}
\addtolength{\abovedisplayshortskip}{\extraspace}
\addtolength{\belowdisplayshortskip}{\extraspace}}
\newcommand{\ee}{\end{equation}}

\newcommand{\ba}{\begin{eqnarray}
\addtolength{\abovedisplayskip}{\extraspaces}
\addtolength{\belowdisplayskip}{\extraspaces}
\addtolength{\abovedisplayshortskip}{\extraspace}
\addtolength{\belowdisplayshortskip}{\extraspace}}
\newcommand{\ea}{\end{eqnarray}}

\newcommand{\nonu}{\nonumber \\[.5mm]}
\newcommand{\A}{&\!\!\!}

\newcommand{\newsection}[1]{
\vspace{7mm} \pagebreak[3] \addtocounter{section}{1}
\setcounter{subsection}{0} \setcounter{footnote}{0}
\begin{center}
{\large {\bf \thesection. #1}}
\end{center}
\nopagebreak
\medskip
\nopagebreak \hspace{3mm}}

\begin{document}

\pagenumbering{arabic}

\begin{center}
{\bf Energy and Momentum of a Stationary Beam of Light in the New
General Relativity}
\end{center}
 \centerline{ Gamal G.L. Nashed and Mohamed M. Mourad$^\ast$}

\bigskip

\centerline{{\it Mathematics Department, Faculty of Science, Ain
Shams University, Cairo, Egypt }}

\centerline{{ \it $^{\ast}$ Mathematics Department, Faculty of
Science, Minia University, Minia, Egypt }}
\bigskip
 \centerline{ e-mail:nasshed@asunet.shams.edu.eg}

\hspace{2cm}
\\
\\
\\
\\
\\
\\
\\
\\
\\

We give an exact solution to the gravitational  field in the new
general relativity. The solution creates Bonnor spacetime. This
spacetime describes the gravitational field of a stationary beam
of light.  The energy and momentum of this solution is calculated
using the energy-momentum complex given by M\o ller in (1978)
within the framework of the Weitzenb{\rm $\ddot{o}$}ck spacetime.

\newpage
\newsection{Introduction}

Since Einstein proposed the general theory of relativity,
relativists have not been able to agree upon a definition of the
energy-momentum complexes associated with the gravitational field
\cite{ MTW}. Bondi \cite{Bh} argued that general relativity does
not permit a non-localizable form of energy, so, in principle we
should expect to be able to find an acceptable definition.

The tetrad formulation of gravitation was considered by M\o ller
in connection with attempts to define the energy of gravitational
field \cite{Mo8,M4}. For a satisfactory description of the total
energy of an isolated system it is necessary that the energy
density of the gravitational field is given in terms of first-
and/or second-order derivatives of the gravitational field
variables. It is well-known that there exists no covariant,
nontrivial expression constructed out of the metric tensor.
However, covariant expressions that contain a quadratic form of
first-order derivatives of the tetrad field are feasible. Thus it
is legitimate to conjecture that the difficulties regarding the
problem of defining the gravitational energy-momentum are related
to the geometrical description of the gravitational field rather
than are an intrinsic drawback of the theory \cite{Mj,MDTC}.

New general relativity (N.G.R.) is a gravitational theory which is
formulated by gauging external (spacetime) translation \cite{HN}
and underlain with the Weitzenb{\rm $\ddot{o}$}ck \cite{HS}
spacetime characterized by the metricity condition and by the
vanishing of curvature tensor. N.G.R with the gravitational
Lagrangian (3) given below describes well all the observed
gravitational phenomena in the same level as the general
relativity (G.R.).  Schwarzschild metric,
Reissner-Nordstr$\ddot{o}$m metric, Weyl metric solutions, Kerr
and Kerr Newman  metric \cite{KT} have been known also in the
N.G.R.

Bonner \cite{Bw} gave an exact solution in the framework of
general relativity theory which describes a stationary beam of
light in the z-direction. Bringley \cite{Bt}  have shown that this
solution is of Kerr-Schild class, then he calculated its
associated  energy using the energy momentum complexes of
Einstein, Landau-Lifshitz, Papapetrou and Weinberg (ELLPW). It is
the aim of the present work
 to derive a solution of an axially symmetric tetrad in the
N.G.R. In section 2 we gave a brief review of the N.G.R. The
axially symmetric tetrad is applied to the field equations of the
N.G.R. in section 3. The solution of the field equations is also
given in section 3. In section 4 we calculated the energy and
momentum  densities associated with the obtained solution using
the definition of the energy-momentum complex given by M\o ller
\cite{Mo8, MWHL}. The final section is devoted to discussion and
main results.
\newsection{The new general relativity theory of gravitation}

In a spacetime with absolute parallelism the parallel vector
fields ${e_i}^\mu$ define the nonsymmetric connection \be
{\Gamma^\lambda}_{\mu \nu} \stackrel{\rm def.}{=} {e_i}^\lambda
{e^i}_{\mu, \nu}, \ee where $e_{i \mu, \nu}=\partial_\nu e_{i
\mu}$. The curvature tensor defined by ${\Gamma^\lambda}_{\mu
\nu}$ is identically vanishing, however.  The metric tensor
$g_{\mu \nu}$
 is given by
 \be g_{\mu \nu}= \eta_{i j} {b^i}_\mu {b^j}_\nu, \ee
with the Minkowski metric $\eta_{i j}=\textrm {diag}(+1\; ,-1\;
,-1\; ,-1)$\footnote{ Latin indices are rasing and lowering with
the aid of $\eta_{i j}$ and $\eta^{ i j}$.}.
 We note that, associated with any tetrad field ${e_i}^\mu$ there
 is a metric field defined
 uniquely by (2), while a given metric $g^{\mu \nu}$ does not
 determine the tetrad field completely; for any local Lorentz
 transformation of the tetrads ${e_i}^\mu$ leads to a new set of
 tetrads which also satisfy (2).
  The gravitational Lagrangian ${\cal L_G}$ has the form \cite{KT}
\be {\cal L}_G  =  \sqrt{-g} L_G=\sqrt{-g} \left(-\displaystyle{1
\over 3 \kappa}(t^{\mu \nu \lambda}t_{\mu \nu \lambda}-  \Phi^\mu
\Phi_\mu)+\xi a^\mu a_\mu\right),\ee where $t_{\mu \nu \lambda}$,
$ \Phi_\mu$ and $a_\mu$ are  irreducible representation of the
torsion tensor defined by \ba t_{\mu \nu \lambda} \A \stackrel{\rm
def.}{=} \A \displaystyle{1 \over 2} (T_{\mu \nu \lambda}+T_{\nu
\mu \lambda})+\displaystyle{1 \over 6}(g_{\lambda
\mu}\Phi_\nu+g_{\lambda \nu}\Phi_\mu)-\displaystyle{1 \over
3}g_{\mu \nu}\Phi_\lambda,\nonu
\Phi_\mu \A \stackrel{\rm def.}{=} \A {T^\lambda}_{\mu
\lambda},\qquad \qquad a_\mu \stackrel{\rm def.}{=}
 \displaystyle{1 \over 6} \epsilon_{\mu \nu \rho \sigma}T_{\nu \rho
\sigma}, \ea with $\epsilon_{\mu \nu \rho \sigma}$ is a totally
antisymmetric tensor normalized to
\[ \epsilon_{0 1 2 3}=-\sqrt{-g}, \qquad \qquad with \qquad \qquad g \stackrel{\rm
def.}{=}  \det.(g_{\mu \nu}),\] and $T_{\mu \nu \lambda}$ is the
torsion tensor defined by
 \be {T^\lambda}_{\mu \nu}\stackrel{\rm
def.}{=}{\Gamma^\lambda}_{\mu \nu}-{\Gamma^\lambda}_{\nu \mu}.\ee
$\kappa$ and $\xi$ are the Einstein gravitational constant and  a
free dimensionless parameter\footnote{Throughout this paper we use
the relativistic units, $c=G=1$ and
 $\kappa=8\pi$.}.

 The gravitational field equations for the
system described by ${\it L_G}$ are the following:

 \be G_{\mu \nu}(\{\})
+H_{\mu \nu} = -{\kappa} T_{\mu \nu}, \ee \be
\partial_\mu\left(J^{i j \mu}\right)=0, \ee where the Einstein tensor
$G_{\mu \nu}(\{\})$ is defined by \be G_{\mu
\nu}(\{\})\stackrel{\rm def.}{=} R_{\mu \nu}(\{\})-{1 \over 2}
g_{\mu \nu} R(\{\}), \ee and $R_{\mu \nu}(\{\})$ is the Ricci
tensor and $R(\{\})$ is the Ricci scalar. We
 assume  that the energy-momentum tensor of matter fields is
symmetric. The energy-momentum tensor of a source field with
Lagrangian $L_M$: \be \sqrt{-g} T^{\mu \nu}  \stackrel{\rm
def.}{=}e^{i \mu} \displaystyle {\delta (-\sqrt{-g} L_M) \over
\delta {e^i}_\nu}.\ee Here $H_{\mu \nu}$ and $J_{i j \mu}$ are
given by \be H^{\mu \nu} \stackrel{\rm def.}{=}
\displaystyle{\kappa \over \lambda} \left[\displaystyle{1 \over 2}
\left\{ \epsilon^{\mu \rho \sigma \lambda}({T^\nu}_{\rho
\sigma}-{T_{\rho \sigma}}^\nu)+\epsilon^{\nu \rho \sigma
\lambda}({T^\mu}_{\rho \sigma}-{T_{\rho
\sigma}}^\mu)\right\}a_\lambda-\displaystyle{3 \over 2}a^\mu
a^\nu-\displaystyle{3 \over 4}g^{\mu \nu}a^\lambda a_\lambda
\right], \ee and

\be J^{i j \mu} \stackrel{\rm def.}{=} -\displaystyle{1 \over
2}{e^i}_\rho {e^j}_\sigma \epsilon^{\rho \sigma \mu \nu} a_\nu,\ee
respectively, where \be \lambda \stackrel{\rm def.}{=}
\displaystyle{4 \over 9} \xi+\displaystyle{1 \over 3\kappa}.\ee

\newsection{An exact solution}
In this section we will seek a solution satisfying the following
conditions: The parallel vector fields having the form \be
{b^k}_\mu={\delta^k}_\mu+M(x,y)l^k l_\mu.\ee Here $M(x,y)$ is a
function of (x,y) and $l_\mu$ is a quantity satisfying the
conditions \be \eta^{\mu \nu}l_\mu l_\nu=0, \quad  and \quad l^k
\quad  is \quad defined \quad by \quad  l^k \stackrel {\rm
def.}{=} {\delta^k}_\mu \eta^{\mu \nu} l_\nu.\ee Applying (13) to
the field equations (6) and (7) one can obtains the values of
$l_\mu$ and $l^k$ in the form \be l_0=1/\sqrt{2} \; , \qquad
l_1=l_2=0\; , \qquad l_3=-1/\sqrt{2}.\ee Writing explicitly the
tetrad (13) using (15) one obtains \ba \A  \A {b^{(0)}}_0 =
1+{M(x,y) \over 2}, \qquad {b^{(0)}}_3={-M(x,y) \over 2},\nonu
 \A \A {b^{(1)}}_1={b^{(2)}}_2 =1 \; ,    \nonu
\A  \A {b^{(3)}}_0=-{b^{(0)}}_3, \qquad {b^{(3)}}_3=1-{M(x,y)
\over 2}  . \ea The metric associated with solution (16) has the
form \be
ds^2=-dx^2-dy^2-[1-M(x,y)]dz^2-2M(x,y)dzdt+[1+M(x,y)]dt^2,\ee
which is the Bonner spacetime that describe a stationary beam of
flow in the z-direction. For solution (16) to satisfy (6) and (7)
then the left hand side of (6) must has the form \be \nabla^2
M(x,y)=16\pi \rho, \quad with \quad
\rho=-{T^3}_3=-{T^0}_3={T^3}_0={T^0}_0.\ee
\newsection{Energy and momentum}

The superpotential is given by \cite{Mo8, MWHL}
  \be {{\cal U}_\mu}^{\nu \lambda} ={(-g)^{1/2} \over
2 \kappa} {P_{\chi \rho \sigma}}^{\tau \nu \lambda}
\left[\Phi^\rho g^{\sigma \chi} g_{\mu \tau}
 -\lambda g_{\tau \mu} \gamma^{\chi \rho \sigma}
-(1-2 \lambda) g_{\tau \mu} \gamma^{\sigma \rho \chi}\right]\; ,
\ee where ${P_{\chi \rho \sigma}}^{\tau \nu \lambda}$ is \be
{P_{\chi \rho \sigma}}^{\tau \nu \lambda} \stackrel{\rm def.}{=}
{{\delta}_\chi}^\tau {g_{\rho \sigma}}^{\nu \lambda}+
{{\delta}_\rho}^\tau {g_{\sigma \chi}}^{\nu \lambda}-
{{\delta}_\sigma}^\tau {g_{\chi \rho}}^{\nu \lambda} \ee with
${g_{\rho \sigma}}^{\nu \lambda}$ being a tensor defined by \be
{g_{\rho \sigma}}^{\nu \lambda} \stackrel{\rm def.}{=}
{\delta_\rho}^\nu {\delta_\sigma}^\lambda- {\delta_\sigma}^\nu
{\delta_\rho}^\lambda. \ee The energy-momentum density is defined
by \cite{Mo8} \be {\tau_\mu}^\nu={{{\cal U}_\mu}^{\nu
\lambda}}_{\; , \ \lambda},\ee where comma denotes ordinary
differentiation.  The energy $E$ contained in a sphere with radius
$R$ is expressed by the volume integral
   \cite{M58} \be
P_\mu(R)=\int_{r=R} \int \int {{{\cal U}_\mu}^{0 \alpha}}_{, \
\alpha} d^3 x=\int_{r=R} \int \int {\tau_\mu}^0 d^3x, \ee with
$P_0(R)=E(R)$ which is the energy and $P_\alpha(R)$ is the spatial
momentum.
 Calculating the necessary components of (22) one can obtains
 \be \tau^{0 0}=\tau^{0 3}=\displaystyle{\nabla^2
M(x,y) \over 16\pi}=\rho,\ee where (18) is used in (24).
\newsection{Main results and discussion}

The tetrad formulation of gravitation was considered by M\o ller
in connection with attempts to define the energy of gravitational
field \cite{Mo8,Mo2}. For a satisfactory description of the total
energy of an isolated system it is necessary that the
energy-density of the gravitational field is given in terms of
first- and/or second-order derivatives of the gravitational field
variables. It is well-known that there exists no covariant,
nontrivial expression constructed out of the metric tensor.
However, covariant expressions that contain a quadratic form of
first-order derivatives of the tetrad field are feasible. Thus it
is legitimate to conjecture that the difficulties regarding the
problem of defining the gravitational energy-momentum are related
to the geometrical description of the gravitational field rather
than are an intrinsic drawback of the theory \cite{Mj,MDTC}.

An exact solution (16) which gives the Bonner metric spacetime has
been given in the new general relativity \cite{KT}. This solution
is axially symmetric and describes a stationary beam of light
flowing in the z-direction. The spacetime of this solution does
not have a singularities at all.

It was shown by M\o ller \cite{Mo26} that the tetrad description
of the gravitational field allows a more satisfactory treatment of
the energy-momentum complex than does general relativity. Unlike
the usual energy momentum complexes in the classical field theory
\cite{LL}, the energy-momentum complex considered above is unique,
in the sense that it does not allow a redefinition, i.e., the
addition of extra quantities, because these quantities would
violate the field equations. Therefore, we have  used the
superpotential (19) to calculate the energy and spatial momentum
densities (22). The energy and momentum densities calculated from
the complex (22) coincide and are equal to the energy and momentum
density components of ${T^\mu}_\nu$. This result is what would be
expected from purely physical arguments.

\end{document}